\documentclass[10pt,a4paper,twocolumn,english,prl,showpacs,superscriptaddress]{revtex4-1}

\usepackage[T1]{fontenc}
\usepackage[latin9]{inputenc}
\usepackage{fancyhdr}
\usepackage{babel}
\usepackage{amsmath}
\usepackage{amsthm}
\usepackage{amssymb}
\usepackage{graphicx}
\usepackage[unicode=true,pdfusetitle,
 bookmarks=true,bookmarksnumbered=false,bookmarksopen=false,
 breaklinks=false,pdfborder={0 0 1},backref=false,colorlinks=false]
 {hyperref}

\makeatletter

\providecommand{\tabularnewline}{\\}

\IfFileExists{candleshape.def}{%
 \input{candleshape.def}}{}
\IfFileExists{dropshape.def}{%
 \input{dropshape.def}}{}
\IfFileExists{TeXshape.def}{%
 \input{TeXshape.def}}{}
\IfFileExists{triangleshapes.def}{%
 \input{triangleshapes.def}}{}

\usepackage{babel}
\usepackage{calrsfs}
\usepackage{hyperref}
\hypersetup{
    colorlinks=true,
    linkcolor=red,
    citecolor=blue,
    filecolor=blue,      
    urlcolor=blue,
}
\makeatother

\begin{document}
\title{\noindent {\normalsize{}Efficiency of a quantum Otto heat engine operating
under a reservoir at effective negative temperatures}}
\author{{\normalsize{}Rogério J. de Assis}}
\address{Instituto de Física, Universidade Federal de Goiás, 74.001-970, Goiânia
- GO, Brazil}
\author{{\normalsize{}Taysa M. de Mendonça}}
\author{{\normalsize{}Celso J. Villas-Boas}}
\address{Departamento de Física, Universidade Federal de São Carlos, 13565-905,
São Carlos, São Paulo, Brazil}
\author{{\normalsize{}Alexandre M. de Souza}}
\author{{\normalsize{}Roberto S. Sarthour}}
\author{{\normalsize{}Ivan S. Oliveira}}
\address{Centro Brasileiro de Pesquisas Físicas, Rua Dr. Xavier Sigaud 150,
22290-180 Rio de Janeiro, Rio de Janeiro, Brazil}
\author{{\normalsize{}Norton G. de Almeida}}
\address{Instituto de Física, Universidade Federal de Goiás, 74.001-970, Goiânia
- GO, Brazil}
\begin{abstract}
{\normalsize{}We perform an experiment in which a quantum heat engine
works under two reservoirs, one at a positive spin temperature and
the other at an effective negative spin temperature}\emph{\normalsize{}
i.e}{\normalsize{}., when the spin system presents population inversion.
We show that the efficiency of this engine can be greater than that
when both reservoirs are at positive temperatures. We also demonstrate
the counter-intuitive result that the Otto efficiency can be beaten
only when the quantum engine is operating in the finite-time mode.}{\normalsize\par}
\end{abstract}
\maketitle
Negative temperature is one of the most exciting current topics in
contemporary physics, being subject to skepticism and criticism in
the scientific community. This topic emerged in 1951 when Purcell
\cite{Purcell1951} first produced spin states with inverted population
and considered the possibility of describing them as states at negative
spin temperatures. In 1956, Ramsey \cite{Ramsey1956} studied these
states theoretically, considering them as states of thermodynamic
equilibrium, and discussed the consequences on the bases of thermodynamics
arising from the incorporation of negative temperatures, one of them
being the need for modifications of thermodynamics laws. After more
than 60 years since the experiment carried out by Purcell \cite{Purcell1951},
another experiment \cite{Carr2013,Braun2013} involving negative temperatures
called attention, which have triggered a discussion on the definition
of equilibrium entropy in statistical mechanics, or the \textquotedblleft Boltzmann
\emph{versus} Gibbs entropy\textquotedblright{} issue \cite{Dunkel2013,Sokolov2013,Hilbert2014,Campisi2015,Cerino2015,Frenkel2015,Swendsen2015,Poulter2016,Swendsen2016,Abraham2017,Hama2018}.
Recently, H. Struchtrup \cite{Struchtrup2018} studied this subject
from a different point of view by considering states at negative temperatures
as nonequilibrium states, referring to them as temperature unstable
states or states with apparent negative temperatures, thus keeping
unchanged the bases of thermodynamics. Although the concept of negative
temperature has proven controversial, fortunately it is not necessary
to enter this debate to investigate a heat engine operating in such
environments. Indeed, both equilibrium and nonequilibrium reservoirs
have already been considered in previous works, see for instance Refs.
\cite{Quan2007,Gemmer2009,Linden2010,Liao2010,Gelbwaser-Klimovsky2013,RoBnagel2014,Alicki2015,Klaers2017,Peterson2018,Tacchino2018}.
However, we will use here the nonequilibrium approach discussed in
Ref. \cite{Struchtrup2018}, referring to states with apparent negative
temperatures as states at effective negative temperatures.

Classical heat engines convert thermal resources into work, which
is maximized for reversible operations in which the entropy production
vanishes. In the quantum realm, both the engine and the reservoirs
can be composed by finite-dimensional systems. Differently from the
classical case, quantum engines can be prepared in physical states
without classical analogues \cite{Tacchino2018}. These quantum states,
in which the working substance as well the reservoirs can be prepared,
give rise to an out-of-equilibrium scenario where it is legitimate
to expect improved heat engines as compared to their classical analogs.
Indeed, it was recently demonstrated that the use of squeezed thermal
reservoirs allows for thermal engines of greater efficiencies \cite{RoBnagel2014,Klaers2017}.
As a special example, the quantum Otto heat engine (QOHE) consists
of two isochoric thermalization branches, one with a cold and another
with a hot thermal reservoir in which the Hamiltonian is fixed, and
two other branches, in which the system is disconnected from the thermal
reservoirs and evolves unitarily \cite{Alicki1979,Kosloff1992,Kieu2004,Quan2007,Gemmer2009,Linden2010,Liao2010,Gelbwaser-Klimovsky2013,Alicki2015,Roulet2018}.
Recently, QOHE operating with reservoirs at positive temperatures
was experimentally performed in the nuclear magnetic resonance (NMR)
context and fully characterized in the finite-time operation mode
\cite{Peterson2018}. This experiment demonstrated that the quantumness
of the work substance is not enough to have gain in efficiency. Furthermore,
the work extracted from a QOHE is limited to the same amount of work
extracted from a classical Otto engine, and besides, the maximum work
is extracted only at the quasi-static operation mode \cite{Peterson2018}.

In this letter we show a proof-of-concept implementation of a QOHE
that operates under a thermal reservoir at a positive spin temperature
and another one at an effective negative spin temperature. As far
as we know, our experiment is the first one to investigate quantum
heat engines with reservoir at negative effective temperatures. As
a consequence of this new approach, we obtained extremely innovative
and counterintuitive results as higher efficiency than the Otto limit
and, for some set of parameters, the faster the process the higher
the efficiency.

\begin{figure}[ptbh]
\centering{}\includegraphics[scale=0.75]{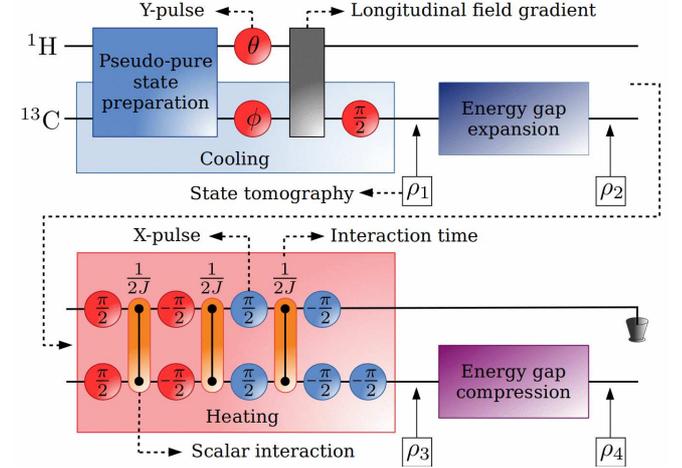}\caption{\label{Fig1} Circuit to QOHE implementation protocol. The blue (red)
circles represent transverse rf-pulses in the $x$ ($y$) direction
that produce rotations by the angles displayed into the circle. The
$\theta$ and $\phi$ angles were adjusted to produce the desired
populations (see the main text).}
\end{figure}

To implement the QOHE we employed a $^{13}\text{C}$-labeled $\text{CHCl}_{3}$
liquid sample diluted in Acetone-D6 and a 500 MHz Varian NMR spectrometer.
Due to dilution, each chloroform molecule ($\text{CHCl}_{3}$) can
be seen as an independent two-qubit system, named $^{13}$C and $^{1}$H
nuclear spins. The coupling interaction between $^{13}$C and $^{1}$H
nucleus is $J=215.1$ Hz and their Larmor frequencies are $\nu_{L}^{H}=500$
MHz and $\nu_{L}^{C}=125$ MHz. As in Ref. \cite{Peterson2018}, the
spin 1/2 of the $^{13}\text{C}$ nucleus is the working medium, and
the spin 1/2 of the $^{1}\text{H}$ nucleus plays the role of the
hot thermal reservoir. The experiments were performed at room temperature.
Radiofrequency pulses allow manipulating the $^{13}$C and $^{1}$H
spins populations separately, and therefore can be used to prepare
different Boltzmann distributions. In the timescale of the experiment
these distributions remain basically unchanged due to the long thermal
relaxation time, which in NMR is associated with the spin lattice
relaxation occurring in a characteristic time $\tau_{1}$ ($\tau_{1}^{H}\sim7,4$
s and $\tau_{1}^{C}\sim11,3$ s). The four-strokes of the quantum
Otto cycle are indicated below and the respective simplified experimental
protocol is shown in Fig. \ref{Fig1}.

\emph{(i) Cooling stroke}. At first, using spatial average techniques
employed by radio-frequency (rf) and gradient fields, the $^{13}\text{\text{C}}$
nuclear spin is prepared in a pseudo-thermal state equivalent to $\rho_{1}=e^{-\beta_{cold}H_{cold}}/Z_{cold}$,
where $\beta_{cold}$ is the cold inverse effective spin temperature,
$H_{cold}$ is the Hamiltonian, and $Z_{cold}$ is the partition function.
The cold inverse effective spin temperature has the form $\beta_{cold}=1/k_{B}T_{cold}$,
with $k_{B}$ being the Boltzmann's constant and $T_{cold}$ the cold
effective spin temperature. The Hamiltonian is given by $H_{cold}=-\frac{1}{2}h\nu_{cold}\sigma_{x}^{C}$,
with $h$ being the Planck's constant, $\nu_{cold}$ a frequency to
be specified, and $\sigma_{x,y,z}^{C}$ the Pauli matrices. 

\emph{(ii) Expansion stroke.} In this stage, from time $t=0$ to $t=\tau$,
the time-modulated rf-field resonant with the $^{13}\text{C}$ nuclear
spin drives the working medium Hamiltonian according to $H_{exp}\left(t\right)=-\frac{1}{2}h\nu\left(t\right)\left[\cos\left(\frac{\pi t}{2\tau}\right)\sigma_{x}^{C}+\sin\left(\frac{\pi t}{2\tau}\right)\sigma_{y}^{C}\right]$,
with $\nu\left(t\right)=\nu_{cold}\left(1-\frac{t}{\tau}\right)+\nu_{hot}\frac{t}{\tau}$,
in a rotating frame at the frequency $\nu_{L}^{C}$. The rf-field
intensity is adjusted so that $\nu_{cold}=2.0$ kHz and $\nu_{hot}=3.6$
kHz, thus expanding the energy gap. The driving time $\tau$ will
be varied into the interval from 100 $\mu$s to 400 $\mu$s. This
time is much shorter than the decoherence scales, which has the order
of seconds, implying that we can describe the driving process as almost
unitary \cite{Peterson2018,Batalhao2014}. Therefore, the \textit{\emph{expansion
stroke}} drives the working medium Hamiltonian to $H_{exp}\left(\tau\right)=-\frac{1}{2}h\nu_{hot}\sigma_{y}^{C}\equiv H_{hot}$
and unitarily evolves the $^{13}\text{C}$ nuclear spin state to $\rho_{2}=U_{\tau,0}\rho_{1}U_{\tau,0}^{\dagger}$,
where $U_{\tau,0}$ stands for the unitary evolution operator.

(\emph{iii) Heating stroke}. Here, the $^{13}\text{C}$ nuclear spin
thermalizes at a hot inverse spin temperature $\beta_{hot}$. The
themalization is achieved by emulating the heat exchange between the
working system and the bath using the $^{1}\text{H}$ nuclear spin
as an auxiliary system, which is previously prepared in a pseudo-thermal
state with inverse spin temperature $\beta_{hot}$. This thermalization
process is effectively achieved by applying a sequence of suitable
rf pulses and free evolution between the nuclei under the scalar interaction
$H_{J}=\frac{1}{4}hJ\sigma_{z}^{C}\sigma_{z}^{H}$, as sketched in
Fig. \ref{Fig1}. At the end of this stage, the $^{13}\text{C}$ nuclear
spin state is in the Gibbs state $\rho_{3}=e^{-\beta_{hot}H_{hot}}/Z_{hot}$.

(\emph{iv) Compression stroke}. At last, this stage is accomplished
by reversing the protocol adopted in the above \emph{expansion stroke},
such that the Hamiltonian is $H_{comp}\left(t\right)=-H_{exp}\left(\tau-t\right)$.
This process is unitary, and at the end the $^{13}\text{C}$ nuclear
spin state is $\rho_{4}=V_{\tau,0}\rho_{3}V_{\tau,0}^{\dagger}$,
where $V_{\tau,0}=U_{\tau,0}^{\dagger}$.

The inverse temperature $\beta_{cold\left(hot\right)}$ of the $^{13}\text{C}$
nuclear spin can be adjusted by means of the population of its excited
state $p_{cold\left(hot\right)}^{+}$ according to the relation

\begin{equation}
\beta_{cold\left(hot\right)}=\frac{1}{h\nu_{cold\left(hot\right)}}\ln\left(\frac{1-p_{cold\left(hot\right)}^{+}}{p_{cold\left(hot\right)}^{+}}\right),\label{eq:Betas}
\end{equation}
where $p_{cold\left(hot\right)}^{+}=\langle+_{cold\left(hot\right)}\vert\rho_{1\left(3\right)}\vert+_{cold\left(hot\right)}\rangle$,
with $\left|+_{cold\left(hot\right)}\right\rangle $ being the eigenstate
of the Hamiltonian $H_{cold(hot)}$ with positive eigenvalue. As can
be seen from Eq. \eqref{eq:Betas}, $p_{cold\left(hot\right)}^{+}\in[0,0.5)$
corresponds to $\beta_{cold\left(hot\right)}$ positive, while $p_{cold\left(hot\right)}^{+}\in(0.5,1]$
corresponds to $\beta_{cold\left(hot\right)}$ negative. In turn,
$p_{cold}^{+}$ is adjusted by rf and gradient fields, as already
mentioned in the cooling stroke\emph{,} and $p_{hot}^{+}$ by adjusting
the population of the excited state of the $^{1}\text{H}$ nuclear
spin, also using rf and gradient fields. In our experiment, see Fig.
\ref{Fig1}, $\phi$ was held fixed, corresponding to $p_{cold}^{+}=0.261\pm0.004$,
whereas $\theta$ was varied so that $p_{hot}^{+}\in(0.5,1]$. The
population $p_{cold\left(hot\right)}^{+}$ is obtained by tomography
of states $\rho_{1\left(3\right)}$ \cite{Leskowitz2004}.

The successive repetition of the procedure \emph{(i)} to \emph{(iv)}
above is equivalent to running successive cycles of the QOHE. Also,
as each experimental realization of the protocol indicated in Fig.
\ref{Fig1} involves spatial averages on a diluted liquid sample containing
about $10^{17}$ noninteracting molecules, each experimental result
presents an average over many copies of a single molecular spin engine
\cite{Abragam1961,Oliveira2007}. All measurements in our experiment
refer to a single realization of the protocol described in Fig. \ref{Fig1}.

It is worthwhile to mention that the finite-time, necessary to accomplish
the expansion and compression stages, is responsible for transitions
between the instantaneous eigenstates of the $^{13}\text{C}$ nuclear
spin Hamiltonian. These transitions result in entropy production,
which introduces irreversibility into the QOHE, causing the poor performance
on thermal engines operating under thermal reservoirs at positive
temperatures \cite{Plastina2014,Cakmak2016,Peterson2018}. Surprisingly,
as shown below, when the QOHE works under one thermal reservoir at
a positive spin temperature and the other at an effective negative
spin temperature, the finite-time operation mode improves the performance
of the QOHE.

To understand our experimental results, firstly we analyze theoretically
the efficiency of the QOHE described previously and, in the following,
we show the results obtained from our experiment. The first quantities
we are interested in are the average net work $\left\langle W\right\rangle $
performed by the QOHE and the average heat $\left\langle Q_{hot}\right\rangle $
absorbed from the hot thermal reservoir, which is all the heat absorbed
by the implemented QOHE.

\begin{figure}[h]
\centering{}%
\begin{tabular}{cc}
\includegraphics[bb=20bp 0bp 564bp 456bp,scale=0.287]{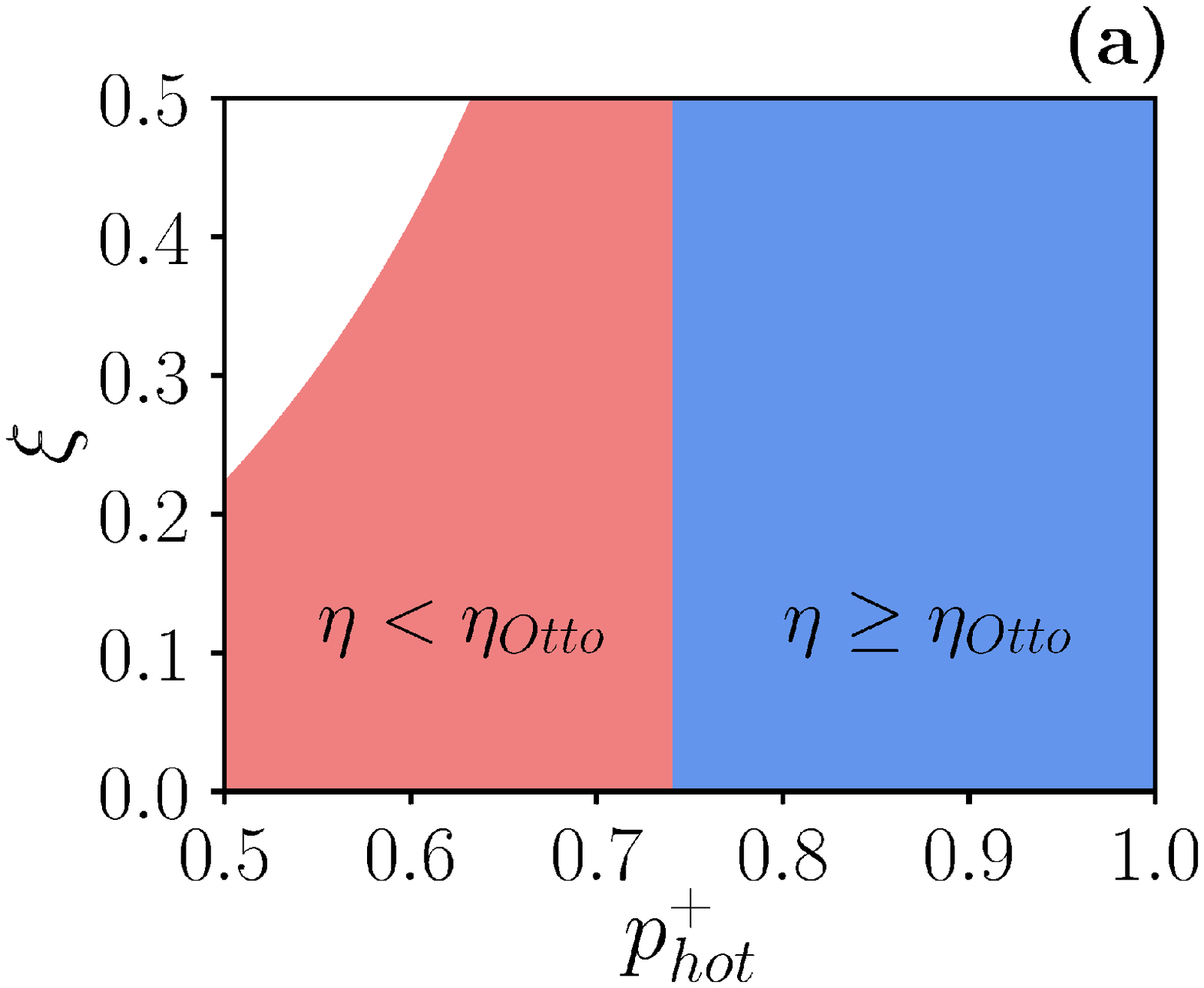} & \includegraphics[bb=110bp -1bp 496bp 441bp,scale=0.287]{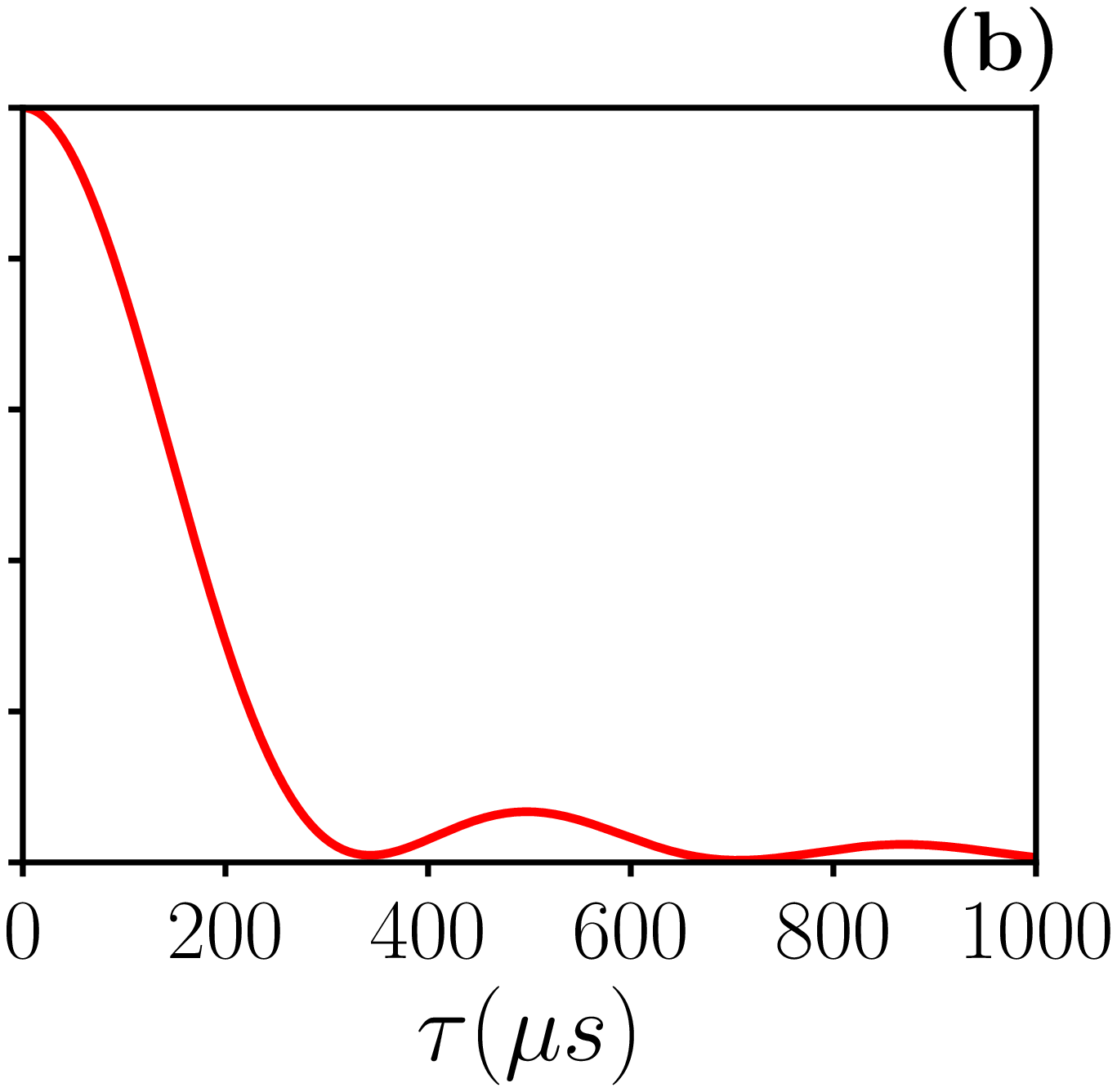}\tabularnewline
\end{tabular}\caption{\label{fig:2}(a) Transition probability $\xi$ versus excited state
population $p_{hot}^{+}$ of the $^{13}\text{C}$ nuclear spin. Red
and blue regions separate the regimes where efficiency is improved
(blue) as compared with the conventional QOHE (red). (b) Transition
probability $\xi$ versus expansion/compression (assumed the same)
time $\tau$. Note that, due to the protocol adopted here, this transition
probability is upper limited by $\xi=1/2$.}
\end{figure}

After a straightforward calculation, using the information contained
in the four-strokes of the QOHE, together with the constraints $\beta_{cold}>0$
and $\beta_{hot}<0$ ($\beta_{hot}=-\left|\beta_{hot}\right|$), we
obtain (see Supplementary Material \cite{SupMat}\nocite{Wang2008})
\begin{multline}
\left\langle W\right\rangle =-\frac{h}{2}\left(\nu_{hot}-\nu_{cold}\right)\left[\tanh\left({\textstyle \frac{1}{2}}\beta_{cold}h\nu_{cold}\right)\right.\\
\left.+\tanh\left({\textstyle \frac{1}{2}}\left|\beta_{hot}\right|h\nu_{hot}\right)\right]+h\xi\left[\nu_{hot}\tanh\left({\textstyle \frac{1}{2}}\beta_{cold}h\nu_{cold}\right)\right.\\
\left.-\nu_{cold}\tanh\left({\textstyle \frac{1}{2}}\left|\beta_{hot}\right|h\nu_{hot}\right)\right]\label{eq:Work}
\end{multline}
and
\begin{gather}
\left\langle Q_{hot}\right\rangle =\frac{h}{2}\nu_{hot}\left[\tanh\left({\textstyle \frac{1}{2}}\beta_{cold}h\nu_{cold}\right)+\tanh\left({\textstyle \frac{1}{2}}\left|\beta_{hot}\right|h\nu_{hot}\right)\right]\nonumber \\
-\xi h\nu_{hot}\tanh\left({\textstyle \frac{1}{2}}\beta_{cold}h\nu_{cold}\right),\label{eq:Heat}
\end{gather}
where $\xi=\left|\langle\pm_{hot}\vert U_{\tau,0}\vert\mp_{cold}\rangle\right|^{2}=\left|\langle\pm_{cold}\vert V_{\tau,0}\vert\mp_{hot}\rangle\right|^{2}$
is the transition probability between the eigenstates $\vert\mp_{cold}\rangle$
and $\vert\pm_{hot}\rangle$. According to our convention, in order
to extract work from the QOHE we must have $\left\langle W\right\rangle <0$.
From Eq. \eqref{eq:Work} this implies
\begin{equation}
\xi<\frac{\left(\nu_{hot}-\nu_{cold}\right)\left[\tanh\left(\frac{1}{2}\beta_{cold}h\nu_{cold}\right)+\tanh\left(\frac{1}{2}\left|\beta_{hot}\right|h\nu_{hot}\right)\right]}{2\left|\nu_{hot}\tanh\left(\frac{1}{2}\beta_{cold}h\nu_{cold}\right)-\nu_{cold}\tanh\left(\frac{1}{2}\left|\beta_{hot}\right|h\nu_{hot}\right)\right|},\label{eq:Condition}
\end{equation}
with $\nu_{hot}\tanh\left(\frac{1}{2}\beta_{cold}h\nu_{cold}\right)-\nu_{cold}\tanh\left(\frac{1}{2}\left|\beta_{hot}\right|h\nu_{hot}\right)\neq0$.
In the case where $\nu_{hot}\tanh\left(\frac{1}{2}\beta_{c}h\nu_{cold}\right)-\nu_{cold}\tanh\left(\frac{1}{2}\left|\beta_{hot}\right|h\nu_{hot}\right)=0$,
as can be seen in Eq. \eqref{eq:Work}, the QOHE performs work regardless
the value of $\xi$. The conditionality to extract work from the QOHE
is graphically shown in the red and blue regions of Fig. \ref{fig:2}
(a). The red region indicates the set of parameters where QOHE operates
as a conventional heat engine, therefore with efficiency $\eta<\eta_{Otto}\equiv1-\nu_{cold}/\nu_{hot}$,
while the blue region, on the other hand, displays the set of parameters
$\xi$ for which efficiency beats that of a conventional QOHE, i.e.,
$\eta\geq\eta_{Otto}$, as will be demonstrated later in the calculations.
Note that there is a blank area in Fig. \ref{fig:2} (a). In that
region the system does not work out as a heat engine. So, not all
values of $\xi$ allows us to have a heat engine for the region $\eta<\eta_{Otto}$.
However, for the region where $\eta\geq\eta_{Otto}$, our system works
out as a heat engine for all values of $\xi$. Fig. \ref{fig:2} (b)
shows that the transition probability goes to zero when the driving
time is increased, as expected by the quantum adiabatic theorem.

Since $\xi$ contains all information about the speed at which the
expansion and compression stages are performed, see Fig. \ref{fig:2}
(b), the contribution to the net work due to the finite-time realization
of these stages lies on the term containing $\xi$ in Eq. \eqref{eq:Work}.
Thus, $\xi$ can be viewed as an adiabaticity parameter. Besides,
this term can be identified with the total inner friction, which is
the difference between the average net work considering actual processes
and the average net work considering ideal (adiabatic) processes \cite{Plastina2014}.
In this way, the total inner friction is related to the nonadiabaticity
of the expansion and compression processes and, therefore, related
to the entropy production \cite{Plastina2014}. When we consider one
of the thermal reservoirs with negative temperature, surprisingly,
$\xi$ may contribute to the increase of the extracted net work, as
can be seen directly in Eq. \eqref{eq:Work}, unlike what happens
when we consider only thermal reservoirs with positive temperatures
\cite{Peterson2018}. If the other QOHE parameters are properly adjusted,
the faster the expansion and compression processes are performed the
greater the contribution of this parameter $\xi$ to the extracted
work, since $\xi$ increases with the shortening of time interval,
see Fig. \ref{fig:2} (b). It is important to note that the increase
of the extracted work with the decrease of time causes the power of
the QOHE to increase, which is a motivating factor for the implementation
of a QOHE operating under effective negative temperature thermal reservoir.
Indeed, this is the main message from Fig.\ref{fig:2} (b).

We can now look at the QOHE efficiency. From Eqs. \eqref{eq:Work}
and \eqref{eq:Heat}, the QOHE efficiency, which is given by $\eta=-\left\langle W\right\rangle /\left\langle Q_{hot}\right\rangle $,
can be written as (see SM)
\begin{equation}
\eta=1-\frac{\nu_{cold}}{\nu_{hot}}\left(\frac{1-\xi{\cal F}}{1-\xi{\cal G}}\right),\label{eq:Efficiency}
\end{equation}
where
\begin{equation}
{\cal F}=\frac{\tanh\left(\frac{1}{2}\left|\beta_{hot}\right|h\nu_{hot}\right)}{\tanh\left(\frac{1}{2}\beta_{cold}h\nu_{cold}\right)+\tanh\left(\frac{1}{2}\left|\beta_{hot}\right|h\nu_{hot}\right)}\label{eq:F}
\end{equation}
 and
\begin{equation}
{\cal G}=\frac{\tanh\left(\frac{1}{2}\beta_{cold}h\nu_{cold}\right)}{\tanh\left(\frac{1}{2}\beta_{cold}h\nu_{cold}\right)+\tanh\left(\frac{1}{2}\left|\beta_{hot}\right|h\nu_{hot}\right)}.\label{eq:G}
\end{equation}
Eq. \eqref{eq:Efficiency} shows that when $\xi=0$ we have $\eta=1-\nu_{cold}/\nu_{hot}\equiv\eta_{Otto}$,
which is the upper limit for the efficiency of Otto cycles operating
under thermal reservoirs both at positive spin temperatures. Strikingly,
as in the case of extracted net work, now $\xi\neq0$ can contribute
to the increase of the QOHE efficiency, causing it to overtake $\eta_{Otto}$.
By analyzing Eqs. \eqref{eq:Efficiency}-\eqref{eq:G}, it is possible
to show that $\beta_{cold}\nu_{cold}<\left|\beta_{hot}\right|\!\nu_{hot}$
implies $\eta<\eta_{Otto}$ while $\beta_{cold}\nu_{cold}\geq\left|\beta_{hot}\right|\!\nu_{hot}$
implies $\eta\geq\eta_{Otto}$, corresponding to the red and blue
regions of Fig. \ref{fig:2}.

\begin{figure}[h]
\centering{}%
\begin{tabular}{cc}
\includegraphics[bb=25bp 0bp 564bp 456bp,scale=0.287]{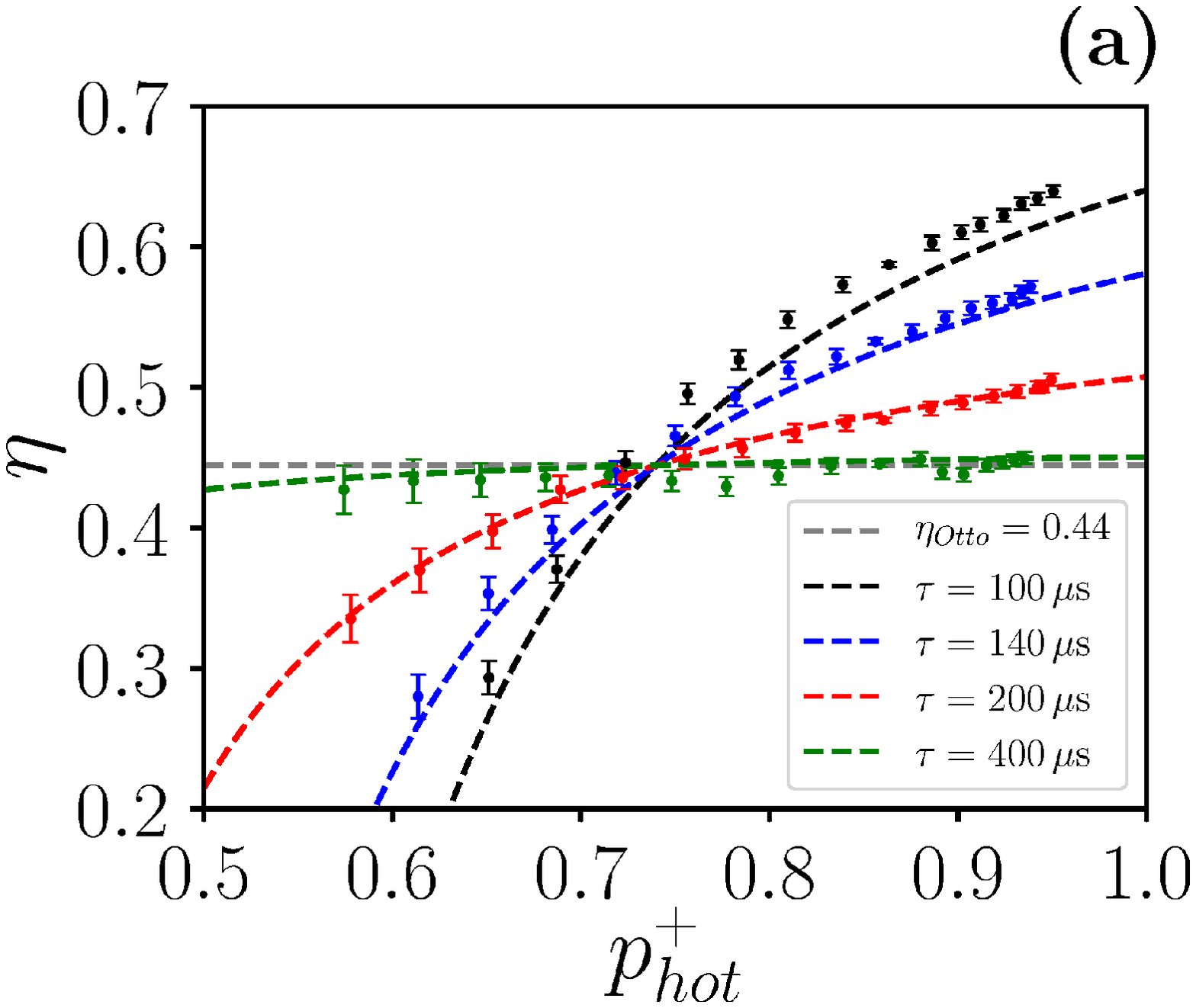} & \includegraphics[bb=110bp 0bp 496bp 441bp,scale=0.287]{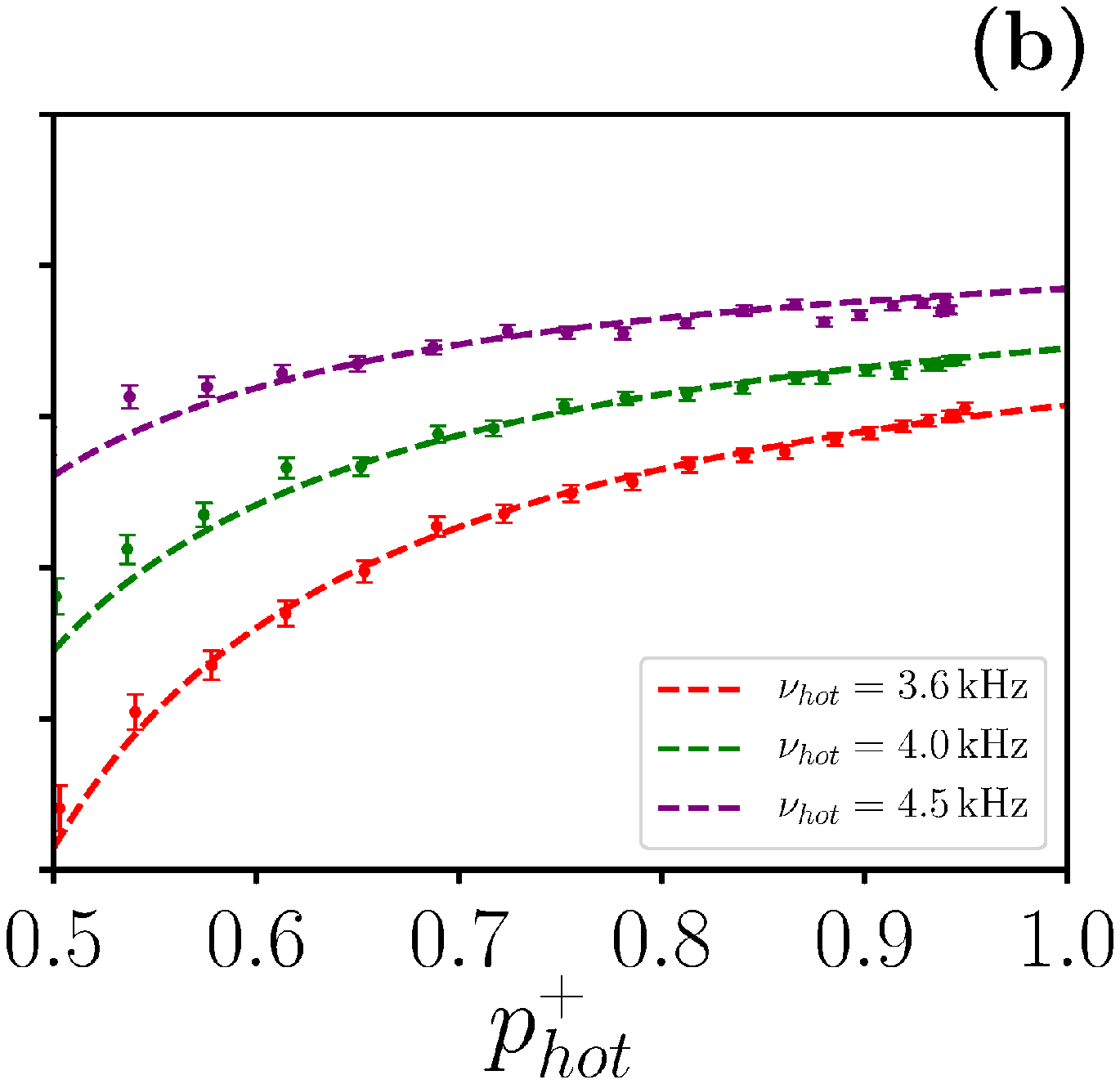}\tabularnewline
\end{tabular}\caption{\label{fig:3} Efficiency ($\eta$) against the population ($p_{hot}^{+}$)
of the excited state. Dashed lines are the theoretical results while
the dots are the experimental measurements. (a) $\eta$ \emph{versu}s
$p_{hot}^{+}$ for several finite operation times, keeping the frequency
ratio fixed ($\nu_{cold}=2.0$ kHz and $\nu_{hot}=3.6$ kHz). The
intersection of the curves happens at the transition point between
the regimes $\eta<\eta_{Otto}$ and $\eta\protect\geq\eta_{Otto}$.
(b) $\eta$ \emph{versu}s $p_{hot}^{+}$ for different frequency ratios
$\nu_{cold}/\nu_{hot}$, with $\nu_{cold}=2$ kHz. The driving time
is fixed in $\tau=200$ $\mu$s.}
\end{figure}

Our experimental results are shown in Figs. \ref{fig:3} (a-b), where
the efficiency of the QOHE is plotted against the population of the
excited state for several driving times $\tau$ ranging from $100$
$\mu$s to $400$ $\mu$s. The efficiency is obtained by means of
the average energy of the $^{13}\text{C}$ nuclear spin after each
stroke, see the Supplementary Material \cite{SupMat}, where the Hamiltonians
$H_{cold}$ and $H_{hot}$ are used together with the nuclear spin
states $\rho_{i}$, $i=1,2,3,4$, obtained by tomography (see Supplementary
Material \cite{SupMat}). Dashed lines are for our theoretical results;
dots are the experimental measurements. Note, in Fig.\ref{fig:3}
(a), the intersection point corresponding to the transition point
from $\eta<\eta_{Otto}$ to $\eta\geq\eta_{Otto}$. This point is
also shown in Fig.\ref{fig:2} (a). It is important to note that the
faster the expansion and compression steps (small driving times $\tau$),
the greater the engine efficiency in the regime in which $\eta\geq\eta_{Otto}$.
In fact, note that to the right of the intersection point, the best
efficiency occurs for $\tau=100$ $\mu$s (black dashed and dots).
Also, since $\eta$ increases with the decreasing of the ratio $\nu_{cold}/\nu_{hot}$,
see Eq. \eqref{eq:Efficiency}, in Fig. \ref{fig:3} (b) we show the
efficiency against the population of the excited state, now for a
fixed driving time $\tau=200$ $\mu$s while varying this frequency
ratio with $\nu_{cold}=2$ kHz. Note that the smaller this ratio the
greater the efficiency, as expected. Therefore, we can use this ratio
to improve the QOHE efficiency.

In conclusion, we experimentally performed a quantum Otto heat engine
(QOHE) in the context of nuclear magnetic resonance by considering
one of the two reservoirs being at effective negative spin temperature.
Such fermionic reservoirs can be engineered, for example, by inverting
the population of a huge nuclear hydrogen spin system. Thus, by weakly
coupling a single nuclear carbon spin to this \textquotedblleft sea\textquotedblright{}
of nuclear spins of the hydrogen atoms, the engineered reservoir is
able to invert the population of the main nuclear spin system (carbon
nuclear spin) as if it were effectively coupled to a negative temperature
reservoir \cite{Mendonca2019}. Unlike previous works with classical
and quantum heat engines, which operate with reservoirs at positive
temperatures, our system provides a set of parameters in which the
faster the processes are performed, the greater the efficiency of
the engine, which we proved experimentally. In this way, our heat
engine is not limited to adiabatic (slow) processes to obtain high
efficiencies. Thus, while the efficiency of conventional Otto engines
reaches the maximum $\eta_{Otto}=1-\nu_{c}/\nu_{h}$ when its expansion
and compression processes occur reversibly, thus in the limit of null
output power, our implemented QOHE reaches $\eta=\eta_{Otto}$ and
$\eta>\eta_{Otto}$ when its expansion and compression processes occur
adiabatically and nonadiabatically, respectively. In addition, for
a QOHE operating under reservoirs at positive spin temperatures only,
the nonadiabaticity as measured by the parameter $\xi=\left|\langle\pm_{hot}\vert U_{\tau,0}\vert\mp_{cold}\rangle\right|^{2}$
comes from the finite-time regime and is responsible to decrease the
engine efficiency. However, in our experiment this parameter $\xi$
can be used to increase the engine efficiency. Besides, our QOHE allows
to obtain efficiency superior to the Otto limit, also proven experimentally,
remembering that we do not take into account the work to engineer
the reservoir at effective negative temperature, following the same
approach of previous works that show superior efficiencies for out
of equilibrium reservoirs, like squeezed ones \cite{RoBnagel2014,Klaers2017}\textit{\emph{.}}
Finally, different from works in Ref. \textit{\emph{\cite{Robnage2016,Lindenfels2018}}}
where the energy is directly converted into mechanical work, our experiment
consists in a proof-of-concept providing the efficiency and maximum
work that can be obtained from our QOHE.  Thus, the results presented
here can trigger new investigations on quantum heat engines in out
of equilibrium reservoirs and applications for effective negative
temperature systems.

\textit{Acknowledgements.} The authors acknowledge useful discussions
and suggestions from Marcelo F. França, Lucas C. Céleri and Daniel
Z. Rossatto. We also acknowledge financial support from the Brazilian
agency, CAPES (Financial code 001) and FAPEG. This work was performed
as part of the Brazilian National Institute of Science and Technology
(INCT) for Quantum Information Grant No. 465469/2014-0. C.J.V.-B.
acknowledges support from Brazilian agencies No. 2013/04162-5 São
Paulo Research Foundation (FAPESP) and from CNPq (Grant No. 308860/2015-2).
A.M.S. acknowledges support from the Brazilian agency FAPERJ (203.166/2017).

\bibliographystyle{apsrev4-1}
\bibliography{IEEEfull,References}

\end{document}


\title{{\LARGE{}Supplementary Material}}
\maketitle

\part{{\LARGE{}Theory}}

\section{The quantum Otto heat engine (QOHE)}

\begin{figure}[H]
\begin{centering}
\includegraphics[scale=0.75]{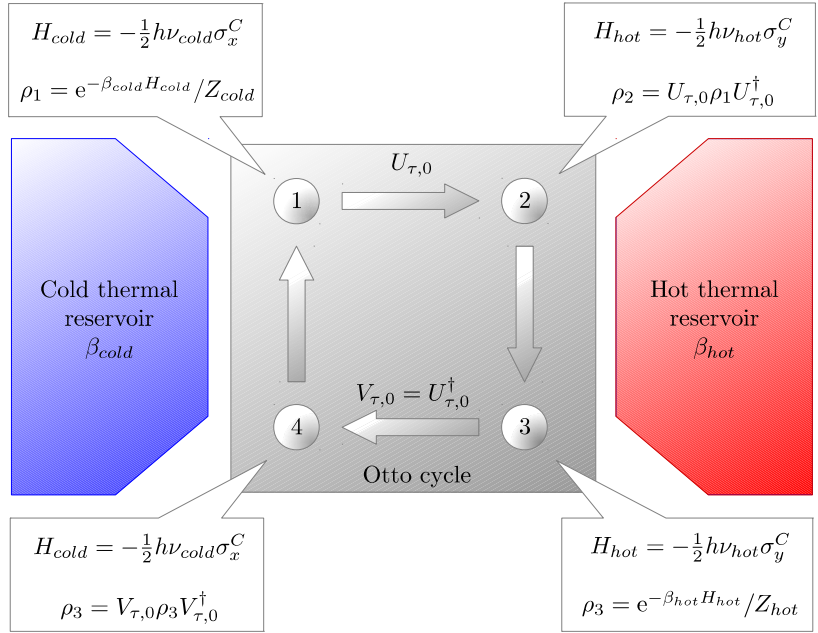} 
\par\end{centering}
\caption{\label{fig:1.1} Scheme of the Otto cycle addressed in the main text,
in which $1\rightarrow2$ corresponds to the \emph{expansion stroke},
$2\rightarrow3$ to the \emph{heating stroke}, $3\rightarrow4$ to
the \emph{compression stroke}, and $4\rightarrow1$ to the \emph{cooling
stroke}. States ($\rho$'s) and Hamiltonians ($H$'s) at the beginning
and end of each stroke are given in the dialog boxes, where $\nu_{cold}$
and $\nu_{hot}$ are frequencies, $\sigma_{x}^{C}$ and $\sigma_{y}^{C}$
are the Pauli matrices related to carbon (the working medium), $\beta_{cold}$
and $\beta_{hot}$ are the inverse temperatures, and $Z_{cold}$ and
$Z_{hot}$ are the partition functions. The unitary operator $U_{\tau,0}$
has the form $U_{\tau,0}={\cal T}\text{e}^{-(i/\hbar)\varint_{0}^{\tau}dtH_{exp}(t)}$,
where ${\cal T}$ is the time-ordering operator and $H_{exp}\left(t\right)=-\frac{1}{2}h\left[\nu_{cold}\left(1-\frac{t}{\tau}\right)+\nu_{hot}\frac{t}{\tau}\right]\left[\cos\left(\frac{\pi t}{2\tau}\right)\sigma_{x}^{C}+\sin\left(\frac{\pi t}{2\tau}\right)\sigma_{y}^{C}\right]$
is the expansion Hamiltonian. The unitary operator $V_{\tau,0}$ can
be obtained through the relation $V_{\tau,0}=U_{\tau,0}^{\dagger}$.}
\end{figure}

The Otto cycle is described in the main text. We thus summarize the
main information about it in Fig. \ref{fig:1.1} such that, when can
refer to it during the calculations.

Analyzing the description of the Otto cycle provided in Fig. \ref{fig:1.1},
we can see that the net work of the cycle is given by

\begin{align}
\left\langle W\right\rangle = & \left\langle W_{1\rightarrow2}\right\rangle +\left\langle W_{3\rightarrow4}\right\rangle \nonumber \\
= & \text{Tr}\left(\rho_{2}H_{hot}\right)-\text{Tr}\left(\rho_{1}H_{cold}\right)+\text{Tr}\left(\rho_{4}H_{cold}\right)-\text{Tr}\left(\rho_{3}H_{hot}\right),\label{eq:1.1}
\end{align}
while the heats exchanged between the working medium and the hot and
cold reservoirs are given by 
\begin{align}
\left\langle Q_{hot}\right\rangle = & \left\langle Q_{3\rightarrow2}\right\rangle \nonumber \\
= & \text{Tr}\left(\rho_{3}H_{hot}\right)-\text{Tr}\left(\rho_{2}H_{hot}\right)\label{eq:1.2}
\end{align}
and 
\begin{align}
\left\langle Q_{cold}\right\rangle = & \left\langle Q_{4\rightarrow1}\right\rangle \nonumber \\
= & \text{Tr}\left(\rho_{1}H_{cold}\right)-\text{Tr}\left(\rho_{4}H_{cold}\right).\label{eq:1.3}
\end{align}
For the Otto cycle to work as a heat engine, the condition $\left\langle W\right\rangle <0$
(work extraction) must be satisfied. Besides, as will be shown below,
the unitary operation of our interest will always provide $\left\langle Q_{hot}\right\rangle >0$
(heat gain) and $\left\langle Q_{cold}\right\rangle <0$ (heat loss).
That said, we can introduce the engine efficiency as being

\begin{equation}
\eta=-\frac{\left\langle W\right\rangle }{\left\langle Q_{hot}\right\rangle }.\label{eq:1.4}
\end{equation}

\section{Calculation of $\left\langle W\right\rangle $, $\left\langle Q_{hot}\right\rangle $,
$\left\langle Q_{cold}\right\rangle $ and $\eta$}

In order to simplify the explicit calculation of the quantities desired,
we will perform separately the traces present in Eqs. (\ref{eq:1.1})-(\ref{eq:1.3}).
In this way we must first compute the partition functions $Z_{cold}$
and $Z_{hot}$. Thus, with the aid of the eigenvalue equation

\begin{equation}
H_{cold\left(hot\right)}\left|\pm_{cold\left(hot\right)}\right\rangle =\pm\frac{1}{2}h\nu_{cold\left(hot\right)}\left|\pm_{cold\left(hot\right)}\right\rangle ,\label{eq:2.1}
\end{equation}
we have 
\begin{align}
Z_{cold}= & \text{Tr}\left(\text{e}^{-\beta_{cold}H_{cold}}\right)\nonumber \\
= & \langle-_{cold}\vert\text{e}^{-\beta_{cold}H_{cold}}\vert-_{cold}\rangle+\langle+_{cold}\vert\text{e}^{-\beta_{cold}H_{cold}}\vert+_{cold}\rangle\nonumber \\
= & \text{e}^{\beta_{cold}h\nu_{cold}/2}+\text{e}^{-\beta_{cold}h\nu_{cold}/2}\nonumber \\
= & 2\cosh\left(\frac{1}{2}\beta_{cold}h\nu_{cold}\right),\label{eq:2.2}
\end{align}
 an similarly, 
\begin{align}
Z_{hot}= & 2\cosh\left(\frac{1}{2}\beta_{hot}h\nu_{hot}\right).\label{eq:2.3}
\end{align}
With this, the calculation of the traces follows as 
\begin{align}
\text{Tr}\left(\rho_{1}H_{cold}\right)= & \frac{1}{Z_{cold}}\text{Tr}\left(\text{e}^{-\beta_{cold}H_{cold}}H_{cold}\right)\nonumber \\
= & \frac{1}{Z_{cold}}\left(\langle-_{cold}\vert\text{e}^{-\beta_{cold}H_{cold}}H_{cold}\vert-_{cold}\rangle+\langle+_{cold}\vert\text{e}^{-\beta_{cold}H_{cold}}H_{cold}\vert+_{cold}\rangle\right)\nonumber \\
= & \frac{h\nu_{cold}}{2Z_{cold}}\left(-\langle-_{cold}\vert\text{e}^{-\beta_{cold}H_{cold}}\vert-_{cold}\rangle+\langle+_{cold}\vert\text{e}^{-\beta_{cold}H_{cold}}\vert+_{cold}\rangle\right)\nonumber \\
= & \frac{h\nu_{cold}}{2Z_{cold}}\left(-\text{e}^{\beta_{cold}h\nu_{cold}/2}+\text{e}^{-\beta_{cold}h\nu_{cold}/2}\right)\nonumber \\
= & -\frac{h\nu_{cold}}{Z_{cold}}\text{sinh}\left(\frac{1}{2}\beta_{cold}h\nu_{cold}\right)\nonumber \\
= & -\frac{1}{2}h\nu_{cold}\text{tanh}\left(\frac{1}{2}\beta_{cold}h\nu_{cold}\right),\label{eq:2.4}
\end{align}
\begin{equation}
\text{Tr}\left(\rho_{3}H_{hot}\right)=-\frac{1}{2}h\nu_{hot}\text{tanh}\left(\frac{1}{2}\beta_{hot}h\nu_{hot}\right),\label{eq:2.5}
\end{equation}
\begin{align}
\text{Tr}\left(\rho_{2}H_{hot}\right)= & \text{Tr}\left(U_{\tau,0}\rho_{1}U_{\tau,0}^{\dagger}H_{hot}\right)\nonumber \\
= & \frac{1}{Z_{cold}}\text{Tr}\left(U_{\tau,0}^{\dagger}H_{hot}U_{\tau,0}\text{e}^{-\beta_{cold}H_{cold}}\right)\nonumber \\
= & \frac{1}{Z_{cold}}\left(\text{e}^{\beta_{cold}h\nu_{cold}/2}\langle-_{cold}\vert U_{\tau,0}^{\dagger}H_{hot}U_{\tau,0}\vert-_{cold}\rangle+\text{e}^{-\beta_{cold}h\nu_{cold}/2}\langle+_{cold}\vert U_{\tau,0}^{\dagger}H_{hot}U_{\tau,0}\vert+_{cold}\rangle\right)\nonumber \\
= & \frac{1}{Z_{cold}}\left[\left(\text{e}^{\beta_{cold}h\nu_{cold}/2}-\text{e}^{-\beta_{cold}h\nu_{cold}/2}\right)\langle-_{cold}\vert U_{\tau,0}^{\dagger}H_{hot}U_{\tau,0}\vert-_{cold}\rangle+\text{e}^{-\beta_{cold}h\nu_{cold}/2}\text{Tr}\left(U_{\tau,0}^{\dagger}H_{hot}U_{\tau,0}\right)\right]\nonumber \\
= & \text{tanh}\left(\frac{1}{2}\beta_{cold}h\nu_{cold}\right)\langle-_{cold}\vert U_{\tau,0}^{\dagger}H_{hot}U_{\tau,0}\vert-_{cold}\rangle-\frac{h\nu_{hot}}{2Z_{cold}}\text{e}^{-\beta_{cold}h\nu_{cold}/2}\text{Tr}\left(\sigma_{y}^{C}\right)\nonumber \\
= & \text{tanh}\left(\frac{1}{2}\beta_{cold}h\nu_{cold}\right)\langle-_{cold}\vert U_{\tau,0}^{\dagger}H_{hot}U_{\tau,0}\vert-_{cold}\rangle\nonumber \\
= & \text{tanh}\left(\frac{1}{2}\beta_{cold}h\nu_{cold}\right)\langle-_{cold}\vert U_{\tau,0}^{\dagger}H_{hot}\left(\vert-_{hot}\rangle\langle-_{hot}\vert+\vert+_{hot}\rangle\langle+_{hot}\vert\right)U_{\tau,0}\vert-_{cold}\rangle\nonumber \\
= & -\frac{1}{2}h\nu_{hot}\text{tanh}\left(\frac{1}{2}\beta_{cold}h\nu_{cold}\right)\langle-_{cold}\vert U_{\tau,0}^{\dagger}\left(\vert-_{hot}\rangle\langle-_{hot}\vert-\vert+_{hot}\rangle\langle+_{hot}\vert\right)U_{\tau,0}\vert-_{cold}\rangle\nonumber \\
= & -\frac{1}{2}h\nu_{hot}\text{tanh}\left(\frac{1}{2}\beta_{cold}h\nu_{cold}\right)\langle-_{cold}\vert U_{\tau,0}^{\dagger}\left(I-2\vert+_{hot}\rangle\langle+_{hot}\vert\right)U_{\tau,0}\vert-_{cold}\rangle\nonumber \\
= & -\frac{1}{2}h\nu_{hot}\text{tanh}\left(\frac{1}{2}\beta_{cold}h\nu_{cold}\right)\left(1-2\left|\langle+_{hot}\vert U_{\tau,0}\vert-_{cold}\rangle\right|^{2}\right)\label{eq:2.6}
\end{align}
and 
\begin{equation}
\text{Tr}\left(\rho_{4}H_{cold}\right)=-\frac{1}{2}h\nu_{cold}\text{tanh}\left(\frac{1}{2}\beta_{hot}h\nu_{hot}\right)\left(1-2\left|\langle+_{cold}\vert V_{\tau,0}\vert-_{hot}\rangle\right|^{2}\right).\label{eq:2.7}
\end{equation}
Calculation details in Eqs. (\ref{eq:2.5}) and (\ref{eq:2.7}), above,
are similar to that in Eqs. (\ref{eq:2.4}) and (\ref{eq:2.6}), respectively,
and therefore were omitted.

Furthermore, we can rewrite Eqs. (\ref{eq:2.6}) and (\ref{eq:2.7})
as 
\begin{equation}
\text{Tr}\left(\rho_{2}H_{hot}\right)=-\frac{1}{2}h\nu_{hot}\text{tanh}\left(\frac{1}{2}\beta_{cold}h\nu_{cold}\right)\left(1-2\xi\right)\label{eq:2.8}
\end{equation}
and 
\begin{equation}
\text{Tr}\left(\rho_{4}H_{cold}\right)=-\frac{1}{2}h\nu_{cold}\text{tanh}\left(\frac{1}{2}\beta_{hot}h\nu_{hot}\right)\left(1-2\xi\right),\label{eq:2.9}
\end{equation}
where 
\begin{equation}
\xi=\left|\langle\pm_{hot}\vert U_{\tau,0}\vert\mp_{cold}\rangle\right|^{2}=\left|\langle\pm_{cold}\vert V_{\tau,0}\vert\mp_{hot}\rangle\right|^{2}.\label{eq:2.10}
\end{equation}
This relation can be demonstrated by observing that 
\begin{align}
\left|\langle+_{cold}\vert V_{\tau,0}\vert-_{hot}\rangle\right|^{2}= & \langle+_{cold}\vert V_{\tau,0}\vert-_{hot}\rangle\langle+_{cold}\vert V_{\tau,0}\vert-_{hot}\rangle^{*}\nonumber \\
= & \langle+_{cold}\vert U_{\tau,0}^{\dagger}\vert-_{hot}\rangle\langle+_{cold}\vert U_{\tau,0}^{\dagger}\vert-_{hot}\rangle^{*}\nonumber \\
= & \langle-_{hot}\vert U_{\tau,0}\vert+_{cold}\rangle^{*}\langle-_{hot}\vert U_{\tau,0}\vert+_{cold}\rangle\nonumber \\
= & \left|\langle-_{hot}\vert U_{\tau,0}\vert+_{cold}\rangle\right|^{2},\label{eq:2.11}
\end{align}
\begin{align}
\left|\langle-_{hot}\vert U_{\tau,0}\vert+_{cold}\rangle\right|^{2}= & \langle-_{hot}\vert U_{\tau,0}\vert+_{cold}\rangle\langle-_{hot}\vert U_{\tau,0}\vert+_{cold}\rangle^{*}\nonumber \\
= & \langle-_{hot}\vert U_{\tau,0}\vert+_{cold}\rangle\langle+_{cold}\vert U_{\tau,0}^{\dagger}\vert-_{hot}\rangle\nonumber \\
= & \langle-_{hot}\vert U_{\tau,0}\left(I-\vert-_{cold}\rangle\langle-_{cold}\vert\right)U_{\tau,0}^{\dagger}\vert-_{hot}\rangle\nonumber \\
= & 1-\langle-_{hot}\vert U_{\tau,0}\vert-_{cold}\rangle\langle-_{cold}\vert U_{\tau,0}^{\dagger}\vert-_{hot}\rangle\nonumber \\
= & 1-\langle-_{cold}\vert U_{\tau,0}^{\dagger}\vert-_{hot}\rangle\langle-_{hot}\vert U_{\tau,0}\vert-_{cold}\rangle\nonumber \\
= & 1-\langle-_{cold}\vert U_{\tau,0}^{\dagger}\left(I-\vert+_{hot}\rangle\langle+_{hot}\vert\right)U_{\tau,0}\vert-_{cold}\rangle\nonumber \\
= & \langle-_{cold}\vert U_{\tau,0}^{\dagger}\vert+_{hot}\rangle\langle+_{hot}\vert U_{\tau,0}\vert-_{cold}\rangle\nonumber \\
= & \langle+_{hot}\vert U_{\tau,0}\vert-_{cold}\rangle\langle+_{hot}\vert U_{\tau,0}\vert-_{cold}\rangle^{*}\nonumber \\
= & \left|\langle+_{hot}\vert U_{\tau,0}\vert-_{cold}\rangle\right|^{2}\label{eq:2.12}
\end{align}
and, similarly, 
\begin{equation}
\left|\langle+_{cold}\vert V_{\tau,0}\vert-_{hot}\rangle\right|^{2}=\left|\langle-_{cold}\vert V_{\tau,0}\vert+_{hot}\rangle\right|^{2}.\label{eq:2.13}
\end{equation}

\noindent As defined, $\xi$ is the transition probability between
the eigenstates of the Hamiltonians $H_{cold}$ and $H_{hot}$. Also,
$\xi$ can be understood as an adiabaticity parameter: when $\xi=0$,
the process obeys the adiabatic theorem and there is no transition
between the instantaneous eigenstates of the Hamiltonian. To positive
temperatures, this correspond to the best possible efficiency. As
soon as $\xi$ becomes non-zero, which implies non-zero power, there
will be transitions which, in the case of positive temperatures, will
lead to a decrease in useful energy and consequently in efficiency
$\eta$. In the case of effective negative temperatures, which is
what we are dealing with, a highly counterintuitive effect occurs,
which is the fact that $\xi$ is non-zero and yet increase the useful
energy to perform work, as we shall demonstrated later.

It is now straightforward to calculate the quantities of interest
$\left\langle W\right\rangle $, $\left\langle Q_{hot}\right\rangle $,
$\left\langle Q_{cold}\right\rangle $ and $\eta$. Using Eqs. (\ref{eq:2.4}),
(\ref{eq:2.5}), (\ref{eq:2.8}) and (\ref{eq:2.9}) in Eqs. (\ref{eq:1.1})-(\ref{eq:1.3}),
we obtain
\begin{multline}
\left\langle W\right\rangle =-\frac{1}{2}h\left(\nu_{hot}-\nu_{cold}\right)\left[\text{tanh}\left(\frac{1}{2}\beta_{cold}h\nu_{cold}\right)-\text{tanh}\left(\frac{1}{2}\beta_{hot}h\nu_{hot}\right)\right]\\
+h\xi\left[\nu_{hot}\text{tanh}\left(\frac{1}{2}\beta_{cold}h\nu_{cold}\right)+\nu_{cold}\text{tanh}\left(\frac{1}{2}\beta_{hot}h\nu_{hot}\right)\right],\label{eq:2.14}
\end{multline}
\begin{equation}
\left\langle Q_{hot}\right\rangle =\frac{1}{2}h\nu_{hot}\left[\text{tanh}\left(\frac{1}{2}\beta_{cold}h\nu_{cold}\right)-\text{tanh}\left(\frac{1}{2}\beta_{hot}h\nu_{hot}\right)\right]-h\xi\nu_{hot}\text{tanh}\left(\frac{1}{2}\beta_{cold}h\nu_{cold}\right)\label{eq:2.15}
\end{equation}
and 
\begin{equation}
\left\langle Q_{cold}\right\rangle =-\frac{1}{2}h\nu_{cold}\left[\text{tanh}\left(\frac{1}{2}\beta_{cold}h\nu_{cold}\right)-\text{tanh}\left(\frac{1}{2}\beta_{hot}h\nu_{hot}\right)\right]-h\xi\nu_{cold}\text{tanh}\left(\frac{1}{2}\beta_{hot}h\nu_{hot}\right).\label{eq:2.16}
\end{equation}
which, when we consider $\beta_{cold}>0$ and $\beta_{hot}<0$ ($\beta_{hot}=-\left|\beta_{hot}\right|$),
can be rewritten as 
\begin{multline}
\left\langle W\right\rangle =-\frac{1}{2}h\left(\nu_{hot}-\nu_{cold}\right)\left[\text{tanh}\left(\frac{1}{2}\beta_{cold}h\nu_{cold}\right)+\text{tanh}\left(\frac{1}{2}\left|\beta_{hot}\right|h\nu_{hot}\right)\right]\\
+h\xi\left[\nu_{hot}\text{tanh}\left(\frac{1}{2}\beta_{cold}h\nu_{cold}\right)-\nu_{cold}\text{tanh}\left(\frac{1}{2}\left|\beta_{hot}\right|h\nu_{hot}\right)\right],\label{eq:2.17}
\end{multline}
\begin{equation}
\left\langle Q_{hot}\right\rangle =\frac{1}{2}h\nu_{hot}\left[\text{tanh}\left(\frac{1}{2}\beta_{cold}h\nu_{cold}\right)+\text{tanh}\left(\frac{1}{2}\left|\beta_{hot}\right|h\nu_{hot}\right)\right]-h\xi\nu_{hot}\text{tanh}\left(\frac{1}{2}\beta_{cold}h\nu_{cold}\right)\label{eq:2.18}
\end{equation}
and 
\begin{equation}
\left\langle Q_{cold}\right\rangle =-\frac{1}{2}h\nu_{cold}\left[\text{tanh}\left(\frac{1}{2}\beta_{cold}h\nu_{cold}\right)+\text{tanh}\left(\frac{1}{2}\left|\beta_{hot}\right|h\nu_{hot}\right)\right]+h\xi\nu_{cold}\text{tanh}\left(\frac{1}{2}\left|\beta_{hot}\right|h\nu_{hot}\right).\label{eq:2.19}
\end{equation}
Finally, when replacing Eqs. (\ref{eq:2.17}) and (\ref{eq:2.18})
in Eq. (\ref{eq:1.4}), we have 
\begin{align}
\eta= & -\frac{\left\langle W\right\rangle }{\left\langle Q_{hot}\right\rangle }\nonumber \\
= & \frac{\frac{1}{2}h\left(\nu_{hot}-\nu_{cold}\right)\left[\text{tanh}\left(\frac{1}{2}\beta_{cold}h\nu_{cold}\right)+\text{tanh}\left(\frac{1}{2}\left|\beta_{hot}\right|h\nu_{hot}\right)\right]-h\xi\left[\nu_{hot}\text{tanh}\left(\frac{1}{2}\beta_{cold}h\nu_{cold}\right)-\nu_{cold}\text{tanh}\left(\frac{1}{2}\left|\beta_{hot}\right|h\nu_{hot}\right)\right]}{\frac{1}{2}h\nu_{hot}\left[\text{tanh}\left(\frac{1}{2}\beta_{cold}h\nu_{cold}\right)+\text{tanh}\left(\frac{1}{2}\left|\beta_{hot}\right|h\nu_{hot}\right)\right]-h\xi\nu_{hot}\text{tanh}\left(\frac{1}{2}\beta_{cold}h\nu_{cold}\right)}\nonumber \\
= & 1-\frac{\nu_{cold}}{\nu_{hot}}\left[\frac{\text{tanh}\left(\frac{1}{2}\beta_{cold}h\nu_{cold}\right)+\text{tanh}\left(\frac{1}{2}\left|\beta_{hot}\right|h\nu_{hot}\right)-2\xi\text{tanh}\left(\frac{1}{2}\left|\beta_{hot}\right|h\nu_{hot}\right)}{\text{tanh}\left(\frac{1}{2}\beta_{cold}h\nu_{cold}\right)+\text{tanh}\left(\frac{1}{2}\left|\beta_{hot}\right|h\nu_{hot}\right)-2\xi\text{tanh}\left(\frac{1}{2}\beta_{cold}h\nu_{cold}\right)}\right]\nonumber \\
= & 1-\frac{\nu_{cold}}{\nu_{hot}}\left\{ \frac{1-2\xi\left[\frac{\text{tanh}\left(\frac{1}{2}\left|\beta_{hot}\right|h\nu_{hot}\right)}{\text{tanh}\left(\frac{1}{2}\beta_{cold}h\nu_{cold}\right)+\text{tanh}\left(\frac{1}{2}\left|\beta_{hot}\right|h\nu_{hot}\right)}\right]}{1-2\xi\left[\frac{\text{tanh}\left(\frac{1}{2}\beta_{cold}h\nu_{cold}\right)}{\text{tanh}\left(\frac{1}{2}\beta_{cold}h\nu_{cold}\right)+\text{tanh}\left(\frac{1}{2}\left|\beta_{hot}\right|h\nu_{hot}\right)}\right]}\right\} \nonumber \\
= & 1-\frac{\nu_{cold}}{\nu_{hot}}\left(\frac{1-2\xi{\cal F}}{1-2\xi{\cal G}}\right)\label{eq:2.20}
\end{align}
where 
\begin{equation}
{\cal F}=\frac{\text{tanh}\left(\frac{1}{2}\left|\beta_{hot}\right|h\nu_{hot}\right)}{\text{tanh}\left(\frac{1}{2}\beta_{cold}h\nu_{cold}\right)+\text{tanh}\left(\frac{1}{2}\left|\beta_{hot}\right|h\nu_{hot}\right)}\label{eq:2.21}
\end{equation}
and 
\begin{equation}
{\cal G}=\frac{\text{tanh}\left(\frac{1}{2}\beta_{cold}h\nu_{cold}\right)}{\text{tanh}\left(\frac{1}{2}\beta_{cold}h\nu_{cold}\right)+\text{tanh}\left(\frac{1}{2}\left|\beta_{hot}\right|h\nu_{hot}\right)}.\label{eq:2.22}
\end{equation}

\section{Conditions for the cycle to work as a heat engine}

As already mentioned, for the above-described cycle to work as a heat
engine, the condition $\left\langle W\right\rangle <0$ must be satisfied.
We recall that this work extraction condition is determined by the
direction of power and heat flow, which, in turn, is given by the
polarization rate of the system two-level steady-state \citep{Gelbwaser-Klimovsky2013}.

Applying this condition to Eq. (\ref{eq:2.17}), we obtain

\begin{equation}
0\leq\xi<\frac{\left(\nu_{hot}-\nu_{cold}\right)\left[\text{tanh}\left(\frac{1}{2}\beta_{cold}h\nu_{cold}\right)+\text{tanh}\left(\frac{1}{2}\left|\beta_{hot}\right|h\nu_{hot}\right)\right]}{2\left|\nu_{hot}\text{tanh}\left(\frac{1}{2}\beta_{cold}h\nu_{cold}\right)-\nu_{cold}\text{tanh}\left(\frac{1}{2}\left|\beta_{hot}\right|h\nu_{hot}\right)\right|},\label{eq:3.1}
\end{equation}
which implies 
\begin{equation}
\nu_{hot}>\nu_{cold}\label{eq:3.2}
\end{equation}
and 
\begin{equation}
\frac{\nu_{cold}}{\nu_{hot}}\neq\frac{\text{tanh}\left(\frac{1}{2}\beta_{cold}h\nu_{cold}\right)}{\text{tanh}\left(\frac{1}{2}\left|\beta_{hot}\right|h\nu_{hot}\right)}.\label{eq:3.3}
\end{equation}
Note that, as shown in Fig. 2(b) of the main text, $0\leq\xi<1/2$,
where the upper limit $1/2$ is due to our choice of the expansion
Hamiltonian ($H_{exp}(t)$) that defines the eigenstates used to calculate
$\xi$. In addition, this upper limit tells us that the working medium
always gains heat from the hot (negative) reservoir and always loses
heat to the cold (positive) reservoir, since $\left\langle Q_{hot}\right\rangle >0$
implies 
\begin{equation}
0\leq\xi<\frac{1}{2}\left[1+\frac{\text{tanh}\left(\frac{1}{2}\left|\beta_{hot}\right|h\nu_{hot}\right)}{\text{tanh}\left(\frac{1}{2}\beta_{cold}h\nu_{cold}\right)}\right]\label{eq:3.4}
\end{equation}
and $\left\langle Q_{cold}\right\rangle <0$ implies 
\begin{equation}
0\leq\xi<\frac{1}{2}\left[1+\frac{\text{tanh}\left(\frac{1}{2}\beta_{cold}h\nu_{cold}\right)}{\text{tanh}\left(\frac{1}{2}\left|\beta_{hot}\right|h\nu_{hot}\right)}\right].\label{eq:3.5}
\end{equation}
Since only the values of $\xi$ for which $\left\langle Q_{cold}\right\rangle <0$
fulfills the condition to have a heat engine, the two conditions above
can be better appreciated in Fig. 2(a) of the main text, where the
blank area indicates that the system does not work out as a heat engine.
On the other hand, as also indicated in Fig.2 (a), there is a wide
range for which our system works out as a heat engine for all values
of $\xi$ into that range and yet $\eta\geq\eta_{Otto}$.

\section{Adiabaticity parameter}

Since the parameter $\xi$ has information about the speed with which
the expansion and compression strokes occur (see Fig. 2(b) of the
main text), we call $\xi$ a \emph{parameter of adiabaticity}. This
nomenclature becomes even clearer when we calculate the work extracted
from the QOHE when its expansion and compression strokes occur adiabatically
(quasi-static). This adiabatic work is given by

\begin{align}
\left\langle W^{ad}\right\rangle = & \left\langle W_{1\rightarrow2}^{ad}\right\rangle +\left\langle W_{3\rightarrow4}^{ad}\right\rangle \nonumber \\
= & \text{Tr}\left(\rho_{2}^{ad}H_{hot}\right)-\text{Tr}\left(\rho_{1}H_{cold}\right)+\text{Tr}\left(\rho_{4}^{ad}H_{cold}\right)-\text{Tr}\left(\rho_{3}H_{hot}\right),\label{eq:4.1}
\end{align}
where 
\begin{equation}
\rho_{2}^{ad}=\frac{\text{e}^{-\beta_{hot}^{ad}H_{hot}}}{Z_{hot}^{ad}}\label{eq:4.2}
\end{equation}
and 
\begin{equation}
\rho_{4}^{ad}=\frac{\text{e}^{-\beta_{cold}^{ad}H_{cold}}}{Z_{cold}^{ad}}.\label{eq:4.3}
\end{equation}
In addition, the adiabaticity conditions $\langle\pm_{cold}\vert\rho_{1}\vert\pm_{cold}\rangle=\langle\pm_{hot}\vert\rho_{2}^{ad}\vert\pm_{hot}\rangle$
and $\langle\pm_{hot}\vert\rho_{3}\vert\pm_{hot}\rangle=\langle\pm_{cold}\vert\rho_{4}^{ad}\vert\pm_{cold}\rangle$
imply 
\begin{equation}
\beta_{cold}\nu_{cold}=\beta_{hot}^{ad}\nu_{hot}\label{eq:4.4}
\end{equation}
and 
\begin{equation}
\beta_{hot}\nu_{hot}=\beta_{cold}^{ad}\nu_{cold},\label{eq:4.5}
\end{equation}
respectively. Using Eqs. (\ref{eq:2.4}) and (\ref{eq:2.5}) in Eq.
(\ref{eq:4.1}), together with Eqs. (\ref{eq:4.4}) and (\ref{eq:4.5}),
we obtain 
\begin{align}
\left\langle W^{ad}\right\rangle = & -\frac{1}{2}h\left[\nu_{hot}\text{tanh}\left(\frac{1}{2}\beta_{hot}^{ad}h\nu_{hot}\right)-\nu_{cold}\text{tanh}\left(\frac{1}{2}\beta_{cold}h\nu_{cold}\right)+\nu_{cold}\text{tanh}\left(\frac{1}{2}\beta_{cold}^{ad}h\nu_{cold}\right)-\nu_{hot}\text{tanh}\left(\frac{1}{2}\beta_{hot}h\nu_{hot}\right)\right]\nonumber \\
= & -\frac{1}{2}h\left[\nu_{hot}\text{tanh}\left(\frac{1}{2}\beta_{cold}h\nu_{cold}\right)-\nu_{cold}\text{tanh}\left(\frac{1}{2}\beta_{cold}h\nu_{cold}\right)+\nu_{cold}\text{tanh}\left(\frac{1}{2}\beta_{hot}h\nu_{hot}\right)-\nu_{hot}\text{tanh}\left(\frac{1}{2}\beta_{hot}h\nu_{hot}\right)\right]\nonumber \\
= & -\frac{1}{2}h\left(\nu_{hot}-\nu_{cold}\right)\left[\text{tanh}\left(\frac{1}{2}\beta_{cold}h\nu_{cold}\right)+\nu_{cold}\text{tanh}\left(\frac{1}{2}\left|\beta_{hot}\right|h\nu_{hot}\right)\right].\label{eq:4.6}
\end{align}
Thus, when we compare Eq. (\ref{eq:4.6}) with Eq. (\ref{eq:2.17}),
we can write 
\begin{equation}
\left\langle W\right\rangle =\left\langle W^{ad}\right\rangle +h\xi\left[\nu_{hot}\text{tanh}\left(\frac{1}{2}\beta_{cold}h\nu_{cold}\right)-\nu_{cold}\text{tanh}\left(\frac{1}{2}\left|\beta_{hot}\right|h\nu_{hot}\right)\right],
\end{equation}
in which the term containing $\xi$ gives the departure from adiabaticity.
Note that this departure from adiabaticity is also present in Eqs.
(\ref{eq:2.18})-(\ref{eq:2.20}).

\section{Effect of $\beta_{hot}<0$ on the efficiency of QOHE}

When the expansion and compression strokes occur adiabatically ($\xi=0$),
for both $\beta_{hot}<0$ and $\beta_{hot}>0$, the efficiency of
the QOHE is given by $\eta=1-\nu_{cold}/\nu_{hot}\equiv\eta_{Otto}$.
When operating with $\beta_{hot}>0$, $\eta_{Otto}$ is the maximum
efficiency of the QOHE, decreasing this value according to the non-adiabaticity
($\xi>0$) of the expansion and compression strokes, as expected.
We can now check what happens to the efficiency of the QOHE when $\beta_{hot}>0$
and $\xi>0$. For this purpose, since Eq. Eq. (\ref{eq:3.1}) is satisfied,
we can analyze the efficiency of the QOHE in cases where $\beta_{cold}\nu_{cold}<\left|\beta_{hot}\right|\nu_{hot}$
and $\beta_{cold}\nu_{cold}\geq\left|\beta_{hot}\right|\nu_{hot}$.
For the first case, which corresponds to the red region of Fig. 2
(a) of the main text, we have $\eta<\eta_{Otto}$, since 
\begin{equation}
\frac{1-2\xi{\cal F}}{1-2\xi{\cal G}}<1\label{eq:5.1}
\end{equation}
implies

\begin{align}
\frac{1-2\xi{\cal F}}{1-2\xi{\cal G}}< & 1\nonumber \\
{\cal G}< & {\cal F}\nonumber \\
\text{tanh}\left(\frac{1}{2}\beta_{cold}h\nu_{cold}\right)< & \text{tanh}\left(\frac{1}{2}\left|\beta_{hot}\right|h\nu_{hot}\right).\label{eq:5.2}
\end{align}
For the second case, which corresponds to the blue region of Fig.
2 (a) of the main text, we have $\eta\geq\eta_{Otto}$, since 
\begin{equation}
\frac{1-2\xi{\cal F}}{1-2\xi{\cal G}}\geq1\label{eq:5.3}
\end{equation}
implies 
\begin{align}
\frac{1-2\xi{\cal F}}{1-2\xi{\cal G}}\geq & 1\nonumber \\
{\cal G}\geq & {\cal F}\nonumber \\
\text{tanh}\left(\frac{1}{2}\beta_{cold}h\nu_{cold}\right)\geq & \text{tanh}\left(\frac{1}{2}\left|\beta_{hot}\right|h\nu_{hot}\right).\label{eq:5.4}
\end{align}
Thus, the QOHE operating with $\beta_{hot}<0$ can overcome $\eta_{Otto}$
when $\xi>0$. 

\part{{\LARGE{}Experiment}}

\section{State tomography}

\begin{figure}[H]
\centering{}\includegraphics[bb=0bp 50bp 960bp 460bp,scale=0.3]{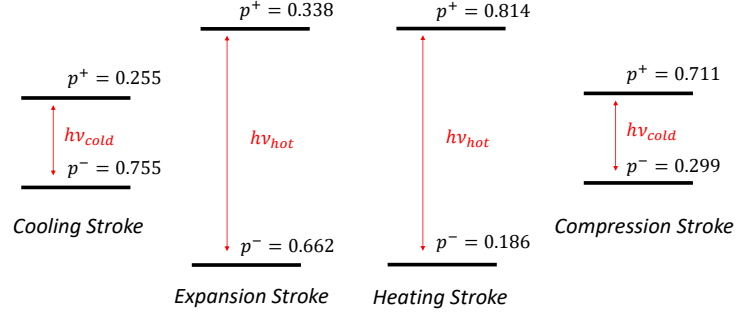}\caption{\label{fig:2} Sketch of the population evolution during the four
cycle stages for $\tau=200$ $\mu$s and the target populations $p_{hot}^{+}=0.813$
and $p_{cold}^{+}=0.26,$ $\nu_{cold}=2.0$ kHz and $\nu_{hot}=3.6$
kHz.}
\end{figure}

\begin{figure}[H]
\begin{centering}
\includegraphics[bb=0bp 0bp 624bp 370bp,scale=0.5]{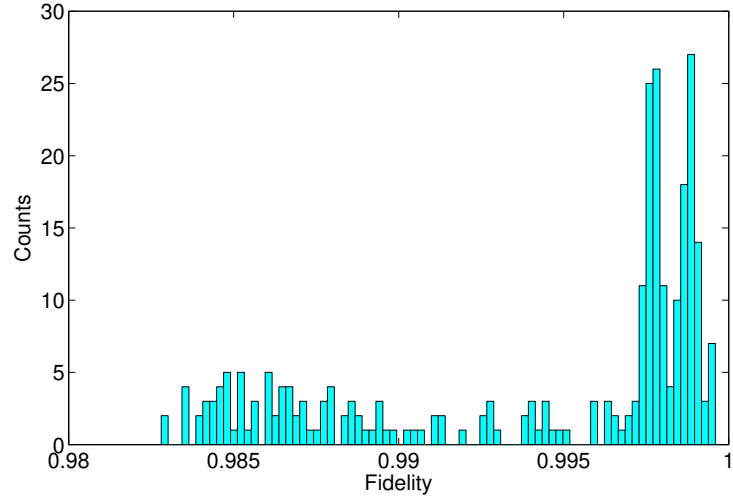}
\par\end{centering}
\caption{\label{fig:3}Quantum fidelity histogram of 444 density matrices obtained
experimentaly. }
\end{figure}

The information required to characterize the quantum Otto cycle was
derived from the $^{13}$C nuclear spin state obtained after each
stroke via quantum state tomography \citep{Leskowitz2004}. To quantify
the ability to implemnt the cycle we have calculated the quantum fidelity
\citep{Wang2008}

\begin{equation}
F=\frac{|Tr(\rho_{exp}\rho_{teo}\dagger)|}{\sqrt{tr(\rho_{exp}^{2})}\sqrt{tr(\rho_{exp}^{2})}}
\end{equation}
between the experimental density matrices ($\rho_{exp})$ and the
theoretical predictions ($\rho_{theo}$) calculated by numerical simulations.
A typical sequence of tomografies is shown in table \ref{tab:1},
for $\tau=200$ $\mu$s and the target populations $p_{hot}^{+}=0.813$
and $p_{cold}^{+}=0.26$, while a sketch ilustrating the population
evolution during the four cycle stages is presented in figure \ref{fig:2}.
In this example, the experimental achieved populations were $p_{hot}^{+}=0.814$
and $p_{cold}^{+}=0.255$, the error associated to the population
preparation was estimated by repeating the state preparation 68 times
and it was found to be approximately $\pm0.004$. Considering all
experiments performed we have the total of 444 tomagraphed matrices,
a histogram of the quantum fidelities for these matrices is shown
in figure \ref{fig:3}. As can be seen, there is a excellent agreement
between the experimental densities matrices and the theoretical matrices,
the obtained fidelities are always above $0.98$, on average the fidelity
is $F=0.9945$. The peak counts with high fidelity values (around
0.998) corresponds to the matrices obtained after the first and second
strokes, while the counts with lower fidelities correspond to the
third and fourth strokes. In the third strokes we emulate the thermalization
of the working system using the $^{1}H$ spin. This thermalization
process is effectively achieved by applying a sequence of suitable
rf pulses and free evolutions, this additional step introduces errors
responsable for the lower fidelities values in the two final strokes.

\begin{table}[H]

\begin{centering}
\begin{tabular}{ccccc}
\toprule 
\textbf{\textsc{\tiny{}Stroke}} & {\tiny{}$\rho{}_{1}$} & {\tiny{}$\rho{}_{2}$} & {\tiny{}$\rho{}_{3}$} & {\tiny{}$\rho{}_{4}$}\tabularnewline
\midrule
\midrule 
\addlinespace
\textbf{\textsc{\tiny{}Experimental}} & {\tiny{}$\left(\begin{array}{cc}
0.47 & 0.24-0.02i\\
0.24+0.02i & 0.53
\end{array}\right)$} & {\tiny{}$\left(\begin{array}{cc}
0.66 & -0.09-0.16i\\
-0.09+0.16i & 0.34
\end{array}\right)$} & {\tiny{}$\left(\begin{array}{cc}
0.51 & -0.01+0.31i\\
-0.01-0.31i & 0.49
\end{array}\right)$} & {\tiny{}$\left(\begin{array}{cc}
0.20 & -0.21-0.10i\\
-0.21+0.10i & 0.70
\end{array}\right)$}\tabularnewline\addlinespace
\midrule 
\addlinespace
\textbf{\textsc{\tiny{}Theory}} & {\tiny{}$\left(\begin{array}{cc}
0.50 & 0.24-0.0i\\
0.24+0.0i & 0.50
\end{array}\right)$} & {\tiny{}$\left(\begin{array}{cc}
0.64 & -0.10-0.17i\\
-0.10+0.17i & 0.36
\end{array}\right)$} & {\tiny{}$\left(\begin{array}{cc}
0.50 & 0.0+0.31i\\
0.0-0.31i & 0.50
\end{array}\right)$} & {\tiny{}$\left(\begin{array}{cc}
0.28 & -0.22-0.03i\\
-0.22+0.03i & 0.72
\end{array}\right)$}\tabularnewline\addlinespace
\midrule 
\textbf{\textsc{\tiny{}Quantum Fidelity}} & {\tiny{}$F=0.9979$} & {\tiny{}$F=0.9988$} & {\tiny{}$F=0.9994$} & {\tiny{}$F=0.9924$}\tabularnewline
\bottomrule
\end{tabular}
\par\end{centering}
\begin{centering}
\caption{\label{tab:1}Experimental end theoretical density matrices for $\tau=200$
$\mu$s and the target populations $p_{hot}^{+}=0.813$ and $p_{cold}^{+}=0.26$. }
\par\end{centering}
\end{table}

\bibliographystyle{plain}
\bibliography{References_SM}